\documentclass[12pt]{article}
\begin{document}
\title{A model for anomalous moisture diffusion through 
a polymer--clay nanocomposite}

\author{A.D. Drozdov\footnote{Corresponding author, fax: +45 9815 3030, 
E--mail: drozdov@iprod.auc.dk}, J. deC. Christiansen\\
{\small Department of Production}\\ 
{\small Aalborg University, Fibigerstraede 16}\\
{\small DK--9220 Aalborg, Denmark}\\
and\\
R.K. Gupta, A.P. Shah\footnote{Present address: Triton Systems Inc., 
200 Turnpike Road, Chelmsford, MA 01824, USA}\\
{\small Department of Chemical Engineering and Constructed Facilities Center}\\
{\small West Virginia University, P.O. Box 6102}\\ 
{\small Morgantown, WV 26506, USA}}
\date{}
\maketitle

\begin{abstract}
Experimental data are reported on moisture diffusion 
and the elastoplastic response of an intercalated nanocomposite 
with vinyl ester resin matrix and montmorillonite clay filler 
at room temperature.
Observations in diffusion tests show that water transport in
the neat resin is Fickian, whereas it becomes 
anomalous (non-Fickian) with the growth of the clay content.
This transition is attributed to immobilization of penetrant
molecules on the surfaces of hydrophilic clay layers.
Observations in uniaxial tensile tests demonstrate that the 
response of vinyl ester resin is strongly elastoplastic,
whereas an increase in the clay content results in a severe
decrease of plastic strains observed as a noticeable reduction 
of curvatures of the stress--strain diagrams.
This is explained by slowing down of molecular
mobility in the host matrix driven by confinement of chains
in galleries between platelets.
Constitutive equations are developed for the anomalous moisture 
diffusion through and the elastoplastic behavior of
a nanocomposite.
Adjustable parameters in these relations are found by fitting 
the experimental data.
Fair agreement is demonstrated between the observations and
the results of numerical simulation.
A striking similarity is revealed between changes in diffusivity,
ultimate water uptake and the rate of plastic flow with an increase
in the clay content.
\end{abstract}
\vspace*{5 mm}

\newpage

\section{Introduction}

This paper is concerned with modelling moisture diffusion through 
a hybrid composite consisting of a polymer matrix filled with clay 
nanoparticles.
Permeation of small molecules through polymer--clay 
nanocomposites has attracted substantial attention in 
the past decade, see, e.g.,
\cite{YUO93,LPP94,BG95,MG95,YUO97,PKS00,Bea01,BMH02,GTV02,TZT02,TVG02,WZQ02},
to mention a few.
This may be explained by a strongly increasing interest to
development and characterization of nanocomposites 
whose physical properties demonstrate a noticeably improvement
compared to those of pristine polymers 
at rather low (conventionally, less than 5\%) filler concentrations.

In the previous works, barrier properties of polymer--clay 
nanocomposites were measured by the permeability cup method
\cite{YUO93,MG95,YUO97,PKS00,TZT02}
and the microgravimetric technique \cite{GTV02,TVG02,WZQ02}.
The former approach characterizes the steady-state mass transport
only.
In the latter case, the microgravimetric analysis focused 
only on (i) the initial parts of the mass gain curves (which provided
estimates of diffusivity for the Fickian diffusion)
and (ii) the equilibrium water uptakes.
This implies that the kinetics of moisture diffusion in polymer--clay
nanocomposites have not yet been investigated in detail.

Changes in diffusivity and permeability of nanocomposites
with clay content are conventionally explained within the
concept of tortuous paths \cite{Nie67}.
According to this approach, the path that a small molecule of 
a penetrant must travel in a polymeric matrix substantially 
increases in the presence of intercalated and/or exfoliated clay 
layers (that possess aspect ratios of order of 10$^{3}$).
This results in a noticeable decrease in the coefficient of diffusion
(estimated as an average value of the square of the end-to-end
distance passed per unit time).
According to the Nielsen equation \cite{Nie67},
the tortuosity factor (the ratio of diffusivity of a composite 
to that of the pristine polymer) is inversely proportional
to a linear function of the volume fraction of filler.
Among other phenomenological relations, it is worth noting
the Cussler formula \cite{CHW88} (the tortuosity factor
is inversely proportional to a linear function of the square 
of the volume fraction of filler),
the Barrer formula \cite{PKS00} (the tortuosity factor
is proportional to the square of the volume fraction of
polymer),
and the power-law equation \cite{FK97} (the tortuosity factor
is proportional to a fractional power of the volume fraction of
matrix).
The Nielsen formula was recently used to fit experimental data 
for polyimide--clay \cite{YUO93,YUO97}
and polyester--clay \cite{BMH02} nanocomposites.
The Barrer equation was employed to match observations
for Nylon 11 filled with nanosize silica \cite{PKS00}.
In all cases, the quality of approximation was far from being
excellent.

To improve the accuracy of fitting, Fredrickson and Bicerano \cite{FB99}
and Bharadwaj \cite{Bha01} proposed phenomenological
relations that account for the orientation of platelets,
their shape, and the state of delamination.
This (merely geometrical) approach may reduce deviations between
experimental data and results of numerical simulation
(due to an increase in the number of adjustable parameters),
but it cannot overcome the main shortcoming of the original 
concept: the tortuous path theory is grounded on the assumption that
the presence of nanoparticles does not affect the diffusivity of
a polymer matrix \cite{FB99}.

This postulate appears to be rather questionable, because 
observations show that mass transport in polymers
is intimately connected with their viscoelastic properties \cite{CMM02}.
Experimental data in oscillation tests with small strains reveal
an increase in the storage modulus and a decrease in the loss
modulus with filler content, $\nu$, for nanocomposites with 
Nylon 6 \cite{GL01}, Nylon 11 \cite{PKS00}, 
Nylon 1012 \cite{WZQ02} and polyester \cite{BMH02} matrices.
The reduction of the loss tangent (that reflects changes in the
viscoelastic response) is accompanied by a decay
in viscoplastic properties revealed as formation 
of a pronounced yield point for polypropylene--clay 
nanocomposites and transition from ductile to brittle fraction 
in Nylon 6--clay nanocomposites with the growth of the filler
fraction \cite{GL01}.
These observations demonstrate that mobility of macromolecules 
in a polymeric matrix slows down with the clay content.
Adopting the concept of ``chain immobilization factor," see 
\cite{FK97} and the references therein, one can conclude that
this deceleration of molecular mobility should be accompanied by
a decrease in diffusivity of small molecules, which is not
accounted for by the concept of tortuous paths.

The objective of this study is three-fold:
\begin{enumerate}
\item
To report experimental data in moisture diffusion tests
and in uniaxial tensile tests on vinyl ester resin--montmorillonite 
clay nanocomposite with various contents of filler.

\item
To derive constitutive equations for water sorption and
stress--strain relations for uniaxial deformation of a
nanocomposite and to find adjustable parameters in
the governing equations by fitting the observations.

\item
To demonstrate that the clay fraction affects the coefficient 
of diffusion and the constants responsible for the viscoplastic 
behavior in a similar way.
\end{enumerate}

The choice of vinyl ester resin for the experimental analysis
is explained by numerous applications of this thermosetting polymer 
as a matrix for glass-reinforced polymer composites 
employed in construction and repair of civil structures \cite{PRT99}.

Montmorillonite (MMT) is an inorganic clay conventionally used
for preparation of hybrid nanocomposites.
It possesses a layered structure constructed of two tetrahedral 
sheets of silica surrounding an octahedral sheet of alumina or magnesia.
The layers (with thickness of 1 nm) are stacked by weak dipole forces,
while the galleries between the layers are occupied by metal cations.

To evaluate the effect of clay content on the mobility of macromolecules
in the matrix, we concentrate of water sorption tests
and uniaxial tensile tests.
Our experimental data (Figures 1 to 5) demonstrate that moisture
diffusion in the neat vinyl ester resin is Fickian.
With an increase in the clay concentration, 
the Fickian diffusion is transformed into an anomalous transport 
of the penetrant molecules.
Anomalous diffusion of water in polymers and polymer composites
has been recently studied in \cite{CW94,VEG99,RXP00,CHL01,US01}.

We suppose that this anomalous moisture uptake in a nanocomposite
filled with MMT particles may be explained by immobilization
of water molecules on the surfaces of hydrophilic clay layers.
A model for a two-stage diffusion with immobilization of 
penetrant molecules was proposed by Carter and Kibler \cite{CK78}
and Gurtin and Yatomi \cite{GY79}.
Unlike these works, we assume that bounded molecules cannot 
leave the sites where they were immobilized, which
allows the number of adjustable parameters to be substantially reduced.
The diffusion process is described by differential equations with
three material constants that are found by matching 
the water uptake curves plotted in Figures 1 to 5.

Our observations in tensile tests (Figures 6 to 10) reveal that
the stress--strain curve for a pristine polymer substantially differs
from a straight line, which means that the mechanical response 
of the neat vinyl ester resin is strongly elastoplastic.
With an increase in the clay content, the stress--strain diagrams
become ``less curved," which is associated with a decay in the
material plasticity.
Similar observations were recently reported for nanocomposites
with Nylon 6, polypropylene \cite{GL01}, 
and poly($\varepsilon$-caprolactone) \cite{GTV02} matrices 
filled with MMT clay.

To describe the elastoplastic behavior, we treat a nanocomposite 
as an equivalent network of chains bridged by junctions 
(entanglements, physical and chemical cross-links, and clay platelets).
The elastic response is attributed to elongations of strands 
in the network, whereas the plastic response is attributed to
sliding of junctions with respect to their reference positions
in the medium (the non-affine deformation of the network).
The decrease in the curvatures of the stress--strain diagrams
with the clay content is ascribed to a decay in mobility
of junctions whose motion is restricted by the presence of 
nanoparticles.
A model is developed to describe constrains imposed 
by the clay platelets on the mobility of chains in the host matrix.
The constitutive equations are determined by three adjustable 
parameters with a transparent physical meaning.
These parameters are found by matching the stress--strain curves 
in uniaxial tensile tests with small strains.

Unlike the previous study \cite{DCG02}, where observations
were reported for a nanocomposite filled with MMT clay 
treated with dimethyl benzyl hydrogenated tallow 
quaternary ammonium chloride (which implied that the clay layers
were not chemicaly bonded to chains in the host matrix), 
the present work focuses on transport of water molecules 
through a nanocomposite filled with MMT clay modified 
with vinyl benzyl ammonium chloride (the latter ensures
that some macromolecules are chemically bonded to the clay 
platelets).

The exposition is organized as follows.
Observations in moisture diffusion tests and uniaxial tensile tests 
are reported in Section 2.
A model for water diffusion through a nanocomposite 
is derived in Section 3
and its material constants are found in Section 4.
Constitutive equations for the elastoplastic response 
of a nanocomposite are developed in Section 5.
Adjustable parameters in the stress--strain relations 
are determined in Section 6.
A brief discussion of our findings is presented in Section 7.
Some concluding remarks are formulated in Section 8.

\section{Experimental procedure}

\subsection{Preparation of samples}

The polymer used was DERAKANE 411-350 epoxy vinyl ester 
resin (Dow Chemical Co.) containing 45 wt.-\% of dissolved styrene. 
To cure the resin at room temperature, it was mixed with 0.5 wt.-\% of 6\% cobalt 
naphthenate catalyst (Sigma Aldrich Co.). 
Additionally, 0.05\% of 99\% N,N dimethyl aniline 
(Lancaster Synthesis, Pelham, NH, USA) was used as 
an accelerator, and 1.5\% of methyl ethyl ketone peroxide 
with 9\% of active oxygen (Sigma Aldrich Co.) as an initiator. 

A sodium montmorillonite clay Cloisite NA (Southern Clay Products Inc.,
Gonzales, TX, USA) was treated with vinyl benzyl ammonium
chloride (Sigma Aldrich Co.).
30 g of Cloisite NA was dispersed in 800 ml of distilled water.
Separately, 6.35 g of vinyl benzyl ammonium chloride (that corresponds 
to 100 meq/100 g of clay) was dissolved in 80 ml of distilled water.
After stirring of the clay dispersion with a mechanical stirrer for
5 h, the vinyl benzyl ammonium chloride solution was added drop by
drop in the dispersion.
The mixture was stirred for another 3 h, and the organo--MMT clay
was filtered using a vacuum filter, dried in an oven at 100 $^{\circ}$C,
and ground in a ball mill.

Neat resin coupons were cast by pouring 
the reaction mixture into Teflon molds with the dimensions
50 $\times$ 12.5 mm and the thickness ranging from 0.2 to 0.6 mm. 
The organically treated clay was added to the liquid resin and manually stirred. 
The mixture was then degassed in a vacuum oven to remove air bubbles. 
Afterwards, the catalyst, initiator and accelerator were added.
The mixture was allowed to cure at room temperature for 24 hours, 
and it was subsequently post-cured in an oven for 3 hours at 90 $^{\circ}$C. 

TEM (transmission electron microscopy) micrographs of samples 
show that an intercalated nanocomposite is produced \cite{SGG02}.
It is worth noting, however, that the thickness of interlayers
between platelets does not increase substantially,
which may be explained by cross-linking between unsaturated
sites on the surfaces of clay layers and vinyl groups of
chains in the host matrix \cite{SGG02,MG94}.

\subsection{Diffusion tests}

Diffusion tests were performed by immersing samples
with rectangular cross-section  
in distilled water at room temperature ($T=25$ $^{\circ}$C). 
The samples had a dry mass ranging from 120 to 400 mg.
They were stored in a controlled humidity chamber, 
and contained 0.05 $\pm$ 0.005 wt.-\% of water 
at the beginning of diffusion tests.
The samples were periodically removed, 
blotted dry with a lint-free tissue, 
weighed and re-immersed into water. 
A typical experiment lasted ten days, 
and, on the first day, readings were taken as frequently 
as every 15 min. 
The balance used had an accuracy of 1 mg, 
and 3 to 5 replicate runs were carried out 
for a given set of condition.

Experimental data are presented in the form of a relative mass
gain, $\Phi$, as a function of a reduced time, $\bar{t}$.
The reduced time, $\bar{t}$, is given by
\begin{equation}
\bar{t}=\frac{t}{4l^{2}},
\end{equation}
where $t$ is time elapsed from the beginning of a test,
and $2l$ is thickness of a sample.
The relative moisture uptake, $\Phi$, reads
\begin{equation}
\Phi(\bar{t})=\frac{R(\bar{t})}{R(\infty)},
\end{equation}
where $R(\bar{t})$ is the mass gain at the reduced time $\bar{t}$
and $R(\infty)$ is the maximal mass gain.

Typical moisture uptake curves are plotted in Figures 1 to 5.
These figures demonstrate that diffusion of water molecules
in the neat resin is Fickian, but it becomes strongly anomalous 
with an increase in the clay content.

\subsection{Mechanical tests}

Dumbbell specimens for mechanical tests were prepared 
in accordance with ASTM D638 specification. 
Tensile tests were performed at ambient temperature by
using a 100 kN Instron machine model 8501 with a cross-head 
speed of 0.254 mm/min. 
This strain rate ensures practically isothermal loading conditions.
The strain was measured independently using a strain gauge 
affixed to the mid-point of a specimen. 
The tensile force was measured by a standard load cell.
The engineering stress, $\sigma$, was determined as the ratio
of the tensile force to the cross-sectional area of a specimen
in the stress-free state.
For any clay concentration, $\nu$, four specimens were tested.
Typical stress--strain diagrams are presented in Figures 6 to 10.
These figures show that the mechanical response of the neat resin
is strongly nonlinear, but the nonlinearity of the stress--strain
curves is reduced with an increase in the clay content.

DSC (differential scanning calorimetry) measurements
show that the glass transition temperature of the neat vinyl
ester resin is about 98 $^{\circ}$C.
It increases with the clay content reaching 112 $^{\circ}$C 
at $\nu=1.0$ wt.-\%, and remains practically constant 
at higher concentrations of filler \cite{SGG02}.
At room temperature, the nanocomposite is far below 
its glass transition point, which implies that we can disregard 
its viscoelastic response and attribute the nonlinearity 
of the stress--strain curves to the elastoplastic behavior.

\section{A model for moisture diffusion}

A sample is treated as a rectilinear plate with thickness $2l$.
We introduce Cartesian coordinates $\{ x,y,z \}$,
where the axis $x$ is perpendicular to the middle plane of the plate,
and the axes $y$ and $z$ are located in the middle plane.
Length and width of the plate substantially exceed its thickness, 
which implies that the moisture concentration depends
on the only coordinate $x$.

After the plate is immersed into water, three processes occur in the
nanocomposite:
\begin{enumerate}
\item
sorption of water molecules to the sample faces from the surrounding,

\item
diffusion of the penetrant into the plate,

\item
adsorption of water molecules on the hydrophilic surfaces of clay layers,
where these molecules become immobilized.
\end{enumerate}

It is conventionally accepted that the rate of sorption in glassy polymers
noticeably exceeds the rate of diffusion \cite{CHL01}, which implies 
that the sorption equilibrium is rapidly established.
The equilibrium condition reads
\begin{equation}
n(t,x)\Bigl |_{x=\pm l}=n^{\circ},
\end{equation}
where $n$ is the moisture concentration at time $t$ at point $x$
[the number of water molecules in the polymeric matrix occupying
a volume with the unit area in the plane $(y,z)$ and the thickness 
$dx$ reads $n(t,x)dx$ ],
and $n^{\circ}$ is the equilibrium moisture concentration in the matrix
on the faces of a sample.

Diffusion of a penetrant through a matrix is described by 
the mass conservation law
\begin{equation}
\frac{\partial n}{\partial t}=\frac{\partial J}{\partial x}-\frac{\partial n_{1}}{\partial t},
\end{equation}
where $n_{1}(t,x)$ denotes the concentration of water molecules 
immobilized at the surfaces of clay layers,
and the mass flux, $J(t,x)$, is given by the Fick formula
\begin{equation}
J=D\frac{\partial n}{\partial x}.
\end{equation}
The constant $D$ in Eq. (5) stands for the coefficient of diffusion
through the polymer (which means that only the mass transport 
through the host matrix is taken into account).
It is assumed, however, that $D$ depends on the clay content:
the presence of clay platelets in the polymeric matrix leads to a decrease 
in the chain mobility that provides a driving force for diffusion.

Adsorption of water molecules on the hydrophilic surfaces of clay layers
is governed by the first-order differential equation
\begin{equation}
\frac{\partial n_{1}}{\partial t}=kn(n_{1}^{\circ}-n_{1}).
\end{equation}
Equation (6) means that the rate of immobilization of small molecules
is proportional to 
(i) the penetrant concentration in the matrix, $n$, 
and (ii) the number of ``unoccupied sites" on the surfaces of clay
platelets, $n_{1}^{\circ}-n_{1}$, where $n_{1}^{\circ}$ stands for 
the total number of sites where water molecules are bounded.
The rate of adsorption, $k$, and the maximal concentration of 
unoccupied sites, $n_{1}^{\circ}$, are assumed to depent
on the clay content, because the presence of nanoparticles
affects mobility of chains in the polymeric matrix, and, as a
consequence, its chemical potential.

The following initial conditions are accepted:
\begin{equation}
n(t,x)\Bigl |_{t=0}=0,
\qquad
n_{1}(t,x)\Bigl |_{t=0}=0,
\end{equation}
which mean that the moisture content in a sample before testing
is disregarded.

The water mass gain (per unit mass of the sample) is given by
\begin{equation}
R(t) =\frac{\kappa}{2l\rho} \int_{-l}^{l} \Bigl [ n(t,x)+n_{1}(t,x) \Bigl ] dx,
\end{equation}
where $\kappa$ is the average mass of a penetrant molecule,
$\rho$ is mass density of the polymeric matrix,
and the functions $n$ and $n_{1}$ obey Eqs. (3) to (7).

Our aim now is to transform the governing equations.
Introducing the dimensionless variables
\[ \bar{x}=\frac{x}{l},
\qquad
c=\frac{n}{n^{\circ}},
\qquad
c_{1}=\frac{n_{1}}{n^{\circ}}
\]
and combining Eqs. (1), (4) and (5), we arrive at the mass flux equation
\begin{equation}
\frac{\partial c}{\partial \bar{t}}=\bar{D}\frac{\partial^{2}c}{\partial \bar{x}^{2}}
-\frac{\partial c_{1}}{\partial \bar{t}},
\end{equation}
where 
\begin{equation}
\bar{D}=4D.
\end{equation}
In the new notation, Eq. (6) reads
\begin{equation}
\frac{\partial c_{1}}{\partial \bar{t}}=Kc(C-c_{1}),
\end{equation}
where 
\begin{equation}
K=4kl^{2}n^{\circ},
\qquad
C=\frac{n_{1}^{\circ}}{n^{\circ}}.
\end{equation}
It follows from Eq. (7) that the initial conditions for Eqs. (9) and (11)
are given by
\begin{equation}
c(0,\bar{x})=0,
\qquad
c_{1}(0,\bar{x})=0.
\end{equation}
Bearing in mind that the functions $c(\bar{t},\bar{x})$ and $c_{1}(\bar{t},\bar{x})$ 
are even functions of $\bar{x}$, we find from Eq. (3) that
\begin{equation}
\frac{\partial c}{\partial \bar{x}}(\bar{t},0)=0,
\qquad
c(\bar{t},1)=1.
\end{equation}
Equation (8) implies that
\begin{equation}
R(\bar{t})=\frac{\kappa}{\rho}  n^{\circ} \int_{0}^{1} \Bigl [ c(\bar{t},\bar{x})+c_{1}(\bar{t},\bar{x})
\Bigr ] d\bar{x}.
\end{equation}
According to Eqs. (9) and (11), with an increase in time, $\bar{t}$,
the functions $c$ and $c_{1}$ approach their steady-state values
\begin{equation}
c(\infty,\bar{x})=1,
\qquad
c_{1}(\infty,\bar{x})=C.
\end{equation}
Equations (15) and (16) result in the formula for the maximal moisture uptake,
\begin{equation}
R(\infty)=\frac{\kappa}{\rho} n^{\circ}(1+C),
\end{equation}
The relative mass gain is found from Eqs. (2), (15) and (17),
\begin{equation}
\Phi(\bar{t})=\frac{1}{1+C}\int_{0}^{1} \Bigl [ c(\bar{t},\bar{x})+c_{1}(\bar{t},\bar{x})
\Bigr ] d\bar{x}.
\end{equation}
Equations (9), (11) and (18) with initial conditions (13) and boundary conditions (14)
are determined by 3 adjustable parameters:
\begin{enumerate}
\item
the reduced diffusivity $\bar{D}$,

\item
the rate of moisture adsorption, $K$, on the hydrophilic
surfaces of clay layers,

\item
the reduced concentration of unoccupied sites, $C$,
on the surfaces of clay platelets. 
\end{enumerate}
These quantities are found by fitting the experimental data in diffusion tests.

\section{Fitting observations in diffusion tests}

We begin with matching the experimental data for the neat vinyl resin
depicted in Figure 1.
In the absence of clay particles, one can set $C=0$.
This equality together with Eqs. (11) and (13) implies that $c_{1}(t,x)=0$.
Bearing in mind this formula, we find from Eq. (9) that the function $c$
obeys the conventional diffusion equation
\begin{equation}
\frac{\partial c}{\partial \bar{t}}=\bar{D}\frac{\partial^{2}c}{\partial \bar{x}^{2}}
\end{equation}
with initial condition (13) and boundary conditions (14).
The initial-boundary problem (13), (14) and (19)
is solved numerically by the finite-difference method 
with an explicit algorithm.
We divide the interval $[0,1]$ into $M$ subintervals 
by points $\bar{x}_{m}=m\Delta x$  $(m=0,1,\ldots,M)$
with $\Delta x=1/M$, 
introduce discrete time $\bar{t}_{n}=n\Delta t$,
and replace Eqs. (13), (14) and (19) by their 
finite-difference approximation
\begin{eqnarray}
c(0,\bar{x}_{m}) &=& 0
\quad
(m=0,1,\ldots,M-1),
\qquad
c(0,\bar{x}_{M})=1,
\nonumber\\
c(\bar{t}_{n+1},\bar{x}_{m}) &=& c(\bar{t}_{n},\bar{x}_{m})+\bar{D}\frac{\Delta t}{\Delta x^{2}}
\Bigl [ c(\bar{t}_{n},\bar{x}_{m+1})-2c(\bar{t}_{n},\bar{x}_{m})
\nonumber\\
&& +c(\bar{t}_{n},\bar{x}_{m-1}) \Bigr ]
\quad
(m=1,\ldots,M-1),
\nonumber\\
c(\bar{t}_{n+1},\bar{x}_{0}) &=& c(\bar{t}_{n+1},\bar{x}_{1}),
\qquad
c(\bar{t}_{n+1},\bar{x}_{M})=1
\qquad
(n=1,2,\ldots).
\end{eqnarray}
The relative mass gain is given by Eq. (18),
where the integral is estimated by means of the Euler formula,
\begin{equation}
\Phi(\bar{t}_{n})=\Delta x\sum_{m=0}^{M-1} c(\bar{t}_{n},\bar{x}_{m}).
\end{equation}
The coefficient $\bar{D}$ is determined by the following procedure.
First, we fix some interval $[D_{\min}, D_{\max}]$, where the ``best-fit" value
of $\bar{D}$ is supposed to be located, and divide this interval
into $I$ sub-intervals by the points $\bar{D}_{i}=D_{\min}+i\Delta D$
($i=0,1,\ldots,I$), where $\Delta D=(D_{\max}-D_{\min})/I$.
For any $\bar{D}_{i}$, Eq. (20) is solved numerically (with the steps
$\Delta x=0.05$ and $\Delta t=0.005$ that guarantee
the stability of the numerical algorithm), and the function $\Phi(\bar{t}_{n})$ is
given by Eq. (21).
The ``best-fit" value of $\bar{D}$ is found from the condition
of minimum of the function
\[
F=\sum_{\bar{t}_{k}} \Bigl [ \Phi_{\exp}(\bar{t}_{k})-\Phi_{\rm num}(\bar{t}_{k})
\Bigr ]^{2},
\]
where the sum is calculated over all point, $\bar{t}_{k}$, depicted in Figure 1,
$\Phi_{\exp}(\bar{t}_{k})$ is the relative mass gain measured in the test,
and $\Phi_{\rm num}(\bar{t}_{k})$ is determined by Eq. (21).
After the ``best-fit" coefficient, $\bar{D}_{i}$, is found, this procedure is
repeated for the new interval $[\bar{D}_{i-1},\bar{D}_{i+1}]$ to ensure
an acceptable accuracy of matching observations.

To show that the numerical results are not affected by our choice of
the steps of integration, $\Delta x$ and $\Delta t$, we solve
Eqs. (20) and (21) numerically with the ``best-fit" parameter $\bar{D}$ 
and the new steps, $\Delta x=0.01$ and $\Delta t=0.0005$.
No difference is visible between the curve plotted in Figure 1 and
the curve obtained by numerical integration of Eqs. (20) and (21)
with the decreased steps in time and the spatial coordinate.

The above procedure is repeated to match observations 
in the tests on 4 specimens with various thicknesses, 
and the average diffusivity (over 4 specimens) and its
standard deviation are determined.

The same numerical algorithm is employed to fit experimental data
on nanocomposites with various contents of clay.
The difference in the treatment of observations consists in the following:
\begin{enumerate}
\item
Eq. (20) is replaced by an appropriate finite-difference approximation 
of Eqs. (9) and (11),

\item
the ``best-fit" values are simultaneously determined for 3 adjustable
parameters, $\bar{D}$, $C$ and $K$.
\end{enumerate}
Figures 1 to 5 demonstrate fair agreement between 
the experimental data and the results of numerical analysis.

It is worth noting that our algorithm for the approximation 
of observations differs from that employed 
in \cite{GTV02,TVG02,SGG02}.
In the previous works, the coefficient of diffusion was determined 
by matching the initial parts of the mass gain diagrams 
(where the function $\Phi(\bar{t})$ is approximately linear),
which did not allow immobilization of water molecules on
the surfaces of clay layers to be accounted for.

The average values of $D$ [given $\bar{D}$, the diffusivity, $D$,
is determined by Eq. (10)], $C$ and $K$ (over 4 tests) 
are plotted versus the clay content in Figures 11 to 13
(the vertical bars stand for the standard deviations).
The dependence of diffusivity, $D$, on the clay content, $\nu$,
is approximated by the phenomenological equation
\begin{equation}
D=D_{0}+D_{1}\exp \Bigl (-\frac{\nu}{\nu_{0}} \Bigr ),
\end{equation}
where $D_{0}$, $D_{1}$ and $\nu_{0}$ are adjustable parameters.
The quantity $\nu_{0}$ is determined by the steepest-descent
method to minimize deviations between the experimental data
and their approximation by Eq. (22).
The coefficients $D_{0}$ and $D_{1}$ are found by the least-squares
algorithm.

The effect of clay content, $\nu$, on the steady concentration, $C$, 
of immobilized water molecules is predicted by the linear function
\begin{equation}
C=C_{1}\nu,
\end{equation}
where the coefficient $C_{1}$ is determined by the least-squares
technique.

To describe the influence of clay content, $\nu$, on
the dimensionless rate, $K$, of water adsorption on the clay platelets, 
we employ the phenomenological relation similar to Eq. (22),
\begin{equation}
\log K=K_{0}+K_{1}\exp \Bigl (-\frac{\nu}{\nu_{0}} \Bigr ),
\end{equation}
where $\log=\log_{10}$.
We use the same value, $\nu_{0}=0.55$, that was found by fitting
observations for the coefficient of diffusion with the help of Eq. (22).
The material parameters, $K_{0}$ and $K_{1}$, in Eq. (24)
are determined by the least-squares method.

It follows from Eq. (17) that the maximal moisture uptake by
the polymeric matrix, $R_{\rm pol}=\kappa n^{\circ}/\rho$, 
reads
\begin{equation}
R_{\rm pol}=\frac{R(\infty)}{1+C}.
\end{equation}
Using measurements of the maximal mass gain, $R(\infty)$,
and the values of $C$ determined by matching observations, 
we calculate $R_{\rm pol}$ from Eq. (25).
This quantity is plotted versus clay content, $\nu$, in Figure 14.
By analogy with Eq. (22), we fit the experimental data by the equation 
\begin{equation}
R_{\rm pol}=R_{0}-R_{1}\exp \Bigl (-\frac{\nu}{\nu_{0}} \Bigr ),
\end{equation}
where the coefficients $R_{0}$ and $R_{1}$ are found
by the least-squares algorithm.

Equations (12) and (17) imply that the maximal moisture 
uptake by the clay particles, $R_{\rm clay}=\kappa n_{1}^{\circ}/\rho$,
is given by
\[
R_{\rm clay}=R(\infty)-R_{\rm pol}.
\]
We determine $R_{\rm clay}$ from this equality and present this
quantity as a function of clay content, $\nu$, in Figure 15.
The experimental data are approximated by the linear function
\begin{equation}
R_{\rm clay}=R_{1}\nu,
\end{equation}
where the constant $R_{1}$ is found by the least-squares technique.

Figures 11 to 15 demonstrate an acceptable agreement between
the observations and their approximations by phenomenological 
relations (22) to (24), (26) and (27).

\section{A model for the elastoplastic response}

A nanocomposite is modelled as an equivalent network of chains 
bridged by junctions. 
Because the viscoelastic behavior of the nanocomposite 
far below the glass transition point is disregarded, 
the network is treated as permanent (which means that 
within the experimental time-scale, chains cannot separate 
from junctions as they are agitated by thermal fluctuations).
The elastoplastic behavior of the network is treated as
sliding of junctions with respect to their reference positions in 
the bulk medium.

For uniaxial deformation, sliding of junctions is determined by a plastic
strain $\epsilon_{\rm p}$.
Adopting the conventional hypothesis that the macro-strain, $\epsilon$,
is transmitted to all chains in the network by surrounding 
macromolecules, we find that
\begin{equation}
\epsilon=\epsilon_{\rm e}+\epsilon_{\rm p},
\end{equation}
where $\epsilon_{\rm e}$ is an elastic strain.
We adopt the mean-field approach, according to which
$\epsilon_{\rm e}$ and $\epsilon_{\rm p}$ are treated 
as average elastic and plastic strains per chain.

The rate of changes in the strain, $\epsilon_{\rm p}$, with time, $t$,
is assumed to be proportional to the rate of changes 
in the macro-strain $\epsilon$,
\begin{equation}
\frac{d\epsilon_{\rm p}}{dt} (t)=\varphi (\epsilon_{\rm e}(t)) 
\frac{d\epsilon}{dt}(t),
\end{equation}
where the coefficient of proportionality, $\varphi$, is a function 
of the elastic strain $\epsilon_{\rm e}$.
The function $\varphi(\epsilon_{\rm e})$
vanishes at the zero elastic strain, $\varphi(0) =0$,
monotonically increases with $\epsilon_{\rm e}$,
and approaches some constant $a\in (0,1)$ at relatively 
large elastic strains.
The constant
\[
a=\lim_{\epsilon_{\rm e}\to \infty} \varphi(\epsilon_{\rm e})
\]
determines the rate of sliding in junctions for a developed plastic flow.

At small strains, a chain is modelled as a linear elastic solid with 
the mechanical energy
\[ 
w=\frac{1}{2}\mu \epsilon_{\rm e}^{2},
\]
where $\mu$ is an average rigidity per chain.
Multiplying the energy, $w$, by the number of chains per unit volume, $n_{0}$,
we find the strain energy density per unit volume of a nanocomposite
\begin{equation}
W=\frac{1}{2}E\epsilon_{\rm e}^{2},
\end{equation}
where $E=\mu n_{0}$ is an elastic modulus.

It is worth noting that $E$ may differ from Young's modulus 
that is conventionally determined as the tangent 
of the angle between the tangent straightline to a stress--strain curve 
at small strains and the horizontal axis.
These two quantities coincide when the elastic strain, $\epsilon_{\rm e}$,
equals the macro-strain, $\epsilon$.
When the stress--strain diagram has a pronounced curvature, 
$E$ exceeds Young's modulus found by the traditional method.

For isothermal uniaxial deformation, the Clausius-Duhem inequality reads
\[
Q(t)=-\frac{dW}{dt}(t)+\sigma(t)\frac{d\epsilon}{dt}(t) \geq 0,
\]
where $Q$ is internal dissipation per unit volume.
Substition of Eqs. (28) to (30) into this equality results in
\[
Q(t)= \Bigl [ \sigma(t)-E\epsilon_{\rm e}(t)
\Bigl (1-\varphi(\epsilon_{\rm e}(t))\Bigr ) \Bigr ] \frac{d\epsilon}{dt}(t)\geq 0.
\]
Assuming the expression in square brackets to vanish,
we arrive at the stress--strain relation
\begin{equation}
\sigma(t) =E\epsilon_{\rm e}(t) \Bigl [1-\varphi(\epsilon_{\rm e}(t))\Bigr ].
\end{equation}
To approximate experimental data, we use the phenomenological equation
\begin{equation}
\varphi(\epsilon_{\rm e})=a \Bigl [ 1-\exp\Bigl (-\frac{\epsilon_{\rm e}}{\varepsilon}\Bigr )
\Bigr ],
\end{equation}
which is determined by two adjustable parameters, $a$ and $\varepsilon>0$.

Constitutive equations (28), (29), (31) and (32) contain 3 adjustable parameters:
\begin{enumerate}
\item
the elastic modulus $E$,

\item
the rate of developed plastic flow $a$,

\item
the strain, $\varepsilon$, that characterizes transition to a steady
plastic flow.
\end{enumerate}
These quantities are found by matching the experimental data depicted in
Figures 6 to 10.

\section{Fitting observations in mechanical tests}

It follows from Eqs. (28), (29), (31) and (32) that in a uniaxial tensile test,
the longitudinal stress, $\sigma$, is given by
\begin{equation}
\sigma(\epsilon) = E(\epsilon-\epsilon_{\rm p})\Bigl \{ 1- a 
\Bigl [1-\exp \Bigl (-\frac{\epsilon-\epsilon_{\rm p}}{\varepsilon}\Bigr )
\Bigr ]\Bigr \},
\end{equation}
where the plastic strain, $\epsilon_{\rm p}$, satisfies the nonlinear 
differential equation
\begin{equation}
\frac{d\epsilon_{\rm p}}{d\epsilon}(\epsilon)
=a \Bigl [1-\exp \Bigl (-\frac{\epsilon-\epsilon_{\rm p}}{\varepsilon}\Bigr )
\Bigr ],
\qquad
\epsilon_{\rm p}(0)=0.
\end{equation}

We begin with matching the experimental data in a test on the neat
vinyl ester resin.
To find the constants, $E$, $a$ and $\varepsilon$,
we fix some intervals $[0,a_{\max}]$ and $[0,\varepsilon_{\max}]$, 
where the ``best-fit" parameters $a$ and $\varepsilon$ are assumed to be located,
and divide these intervals into $I$ subintervals by
the points $a_{i}=i\Delta a$ and $\varepsilon_{j}=j\Delta \varepsilon$  ($i,j=1,\ldots,I$)
with $\Delta a=a_{\max}/I$ and $\Delta \varepsilon=\varepsilon_{\max}/I$.
For any pair, $\{ a_{i}, \varepsilon_{j} \}$, 
Eq. (34) is integrated numerically by the Runge--Kutta method
with the step $\Delta \epsilon=1.0\cdot 10^{-5}$.
Given $\{ a_{i}, \varepsilon_{j} \}$, the elastic modulus $E=E(i,j)$ 
is found by the least-squares technique from the condition 
of minimum of the function
\[
F=\sum_{\epsilon_{k}} \Bigl [ \sigma_{\rm exp}(\epsilon_{k})
-\sigma_{\rm num}(\epsilon_{k}) \Bigr ]^{2},
\]
where the sum is calculated over all experimental points,
$\epsilon_{k}$, depicted in Figure 6, 
$\sigma_{\rm exp}$ is the stress measured in a tensile test, 
and $\sigma_{\rm num}$ is given by Eq. (33).
The ``best-fit" parameters $a$ and $\varepsilon$ minimize the function
$F$ on the set $ \{ a_{i}, \varepsilon_{j} \quad (i,j=1,\ldots, I)  \}$.
Fitting the observations results in $\varepsilon=2.11\cdot 10^{-3}$.

We fix this value of $\varepsilon$ and proceed with matching 
observations in tensile tests on other specimens by using only
two material constants, $E$ and $a$, which are determined by the 
above algorithm.
Figures 6 to 10 demonstrate fair agreement between the observations
and the results of numerical simulation.

The average values of $E$ and $a$ (over 4 samples for any 
concentration of clay) are depicted in Figures 16 and 17, 
where the vertical bars stand for the standard deviations.
Figure 16 shows that $E$ is practically independent
of the clay content.

The effect of clay content, $\nu$, on the rate of developed plastic flow,
$a$, is described by the phenomenological equation
\begin{equation}
a=a_{0}+a_{1}\exp \Bigl (-\frac{\nu}{\nu_{0}} \Bigr ),
\end{equation}
where $\nu_{0}$ is found by matching experimental data for the
coefficient of diffusion, $D$, and the coefficions, $a_{0}$ and $a_{1}$ 
are determined by the least-squares method.
Figure 17 demonstrates good agreement between the observations
and their approximation by Eq. (35).

\section{Discussion}

Changes in the elastic modulus of a hybrid nanocomposite
may be explained by the influence of two factors.
On the one hand, it is conventionally accepted that
reinforcement of  a polymeric matrix by rigid particles 
results in an increase in the elastic modulus of the composite.
On the other hand, clay platelets in a melt-intercalated nanocomposite
screen the macro-strain, which is not transmitted to chains in the
close vicinity of nanoparticles (in particular, to chains 
intercalated into the galleries between silica sheets).
As a result, meso-regions of ``occluded" polymer arise similar to
the domains of occluded rubber in particle-reinforced elastomers 
\cite{WRC93}.
These occluded regions are not deformed, which implies that
the total mechanical energy of a nanocomposite (per unit volume) 
and, as a consequence, its elastic modulus, $E$, decrease
with $\nu$.
Our observation that the elastic modulus, $E$, 
is independent of the clay content, $\nu$, (Figure 16)
may be ascribed to a cumulative effect of these two 
micro-mechanisms.

Weak changes in the elastic modulus of intercalated 
nanocomposites and even its decrease with clay content 
were observed by several authors.
Bharadwaj et al. \cite{BMH02} and Xu et al. \cite{XBH02}
reported a decrease in the elastic modulus of nanocomposites
with polyester and epoxy resin matrices, respectively,
and associated this behavior with a strong decay in the degree
of cross-linking induced by the presence of clay particles
\cite{BMH02}.
Lepoittevin et al. \cite{LDP02} observed a decrease in Young's
modulus for poly($\varepsilon$-caprolactone) filled with 
a non-modified MMT clay and a propounced
growth of the elastic modulus of the nanocomposites with
organo-modified clay particles, but did not provide explanations 
for their findings.
Kim et al. \cite{KLH01} suggested that changes in elastic moduli
and toughness of nanocomposites may be attributed to
nucleation of micro-voids in the vicinity of stacks of silicate sheets
that are thought of as initiation sites for cavitation.
This indicates that physical mechanisms governing the effect 
of clay concentration on elastic moduli of hyrid nanocomposites
remain a subject of debate.

Figure 17 demonstrates that the rate of developed plastic flow, $a$,
monotonically decreases with the clay content, $\nu$.
This trend is in agreement with a conventional 
standpoint \cite{HJG02} that the presence of nanoparticles 
in an intercalated nanocomposite leads to reduction of 
segmental mobility of polymeric chains confined 
to galleries between clay platelets .
It is also confirmed by DSC measurements \cite{SGG02}
that show a pronounced increase in the glass transition
temperature, $T_{\rm g}$,  with the clay content
(the growth of $T_{\rm g}$ confirms a decrease in molecular
mobility of the host matrix).

Figure 11 reveals a similar decrease in the rate of transport of moisture
molecules through the nanocomposite.
Fair approximation of the experimental data depicted
in Figures 11 and 17 by the exponential functions 
with a fixed constant, $\nu_{0}$,
implies that the same micro-mechanism may be responsible
for the effect of clay content on the quantities $D$ and $a$.
According to the free volume concept \cite{CT59,Lit86},
an elementary mass-transport event is rationalized as follows.
At some instant, $t$, a small penetrant molecule in a 
host polymer is at rest (thermal oscillations are disregarded)
waiting for creation of a micro-hole nearby.
When a hole with a sufficient size is formed 
(due to thermal motion of chains),
the penetrant molecule hops into the hole, and remains in
the new position until a new hole arises in its neighborhood.
Although this picture is oversimplified (such important factors
are neglected as clustering of water molecules, their weak 
bonding to polymeric chains, obstacles on the path of a penetrant
molecule caused by the presence of inclusions), 
it provides a clear indication that mass transport is
strongly influenced by mobility of chains.
Resemblance of the graphs presented in Figures 11 and 17 
implies that moisture diffusion and elastoplasticity of a 
nanocomposite may reflect the same phenomena at the micro-level 
associated with molecular mobility of the polymeric matrix.

According to the above scenario of the mass-transport process
as a sequence of hopping and waiting events,
slowing down of water diffusion (observed as a
pronounced decrease in diffusivity $D$) 
implies that the waiting times strongly increase.
As a consequence, the probability of aggregation of 
penetrant molecules grows, which results in formation of
small clusters of water in a nanocomposite.
Clustering of small molecules in solid polymers was observed 
in several studies, see \cite{XGN89,SKS95,SMY98} and 
the references therein.

According to the free volume theory, clustering of water
molecules implies an increase in the total mass uptake
by the polymeric matrix (because penetrant molecules 
weakly bounded into aggregates occupy less volume
than individual molecules).
This conclusion is confirmed by the experimental data
for $R_{\rm pol}$ plotted in Figure 14.
It follows from phenomenological equations (22) and (26) 
that an increase in the moisture uptake with clay content
strongly correlates with a decrease in diffusivity (the same
parameter, $\nu_{0}$, is used to fit the observations 
depicted in Figures 11 and 14).

The coefficient of diffusion, $D$, may serve as an average
(over a macro-volume) rate of molecular transport.
The local rate of moisture diffusion can substantially
differ in the close vicinities of the clay particles
(where molecular mobility is severely reduced due to
intercalation of chains into galleries between platelets) 
and in the bulk (where mobility of chains is weakly affected
by the presence of filler).
This implies that the mass transport in a nanocomposite
becomes substantially heterogeneous.

According to the concept of water clustering,
noticeable slowing down of diffusion near inclusions 
implies that clusters of penetrant molecules are mainly 
formed in neighborhoods of clay particles.
This means that immobilization of a water molecule on
the hydrophilic surface of a clay layer may occur in two different ways:
\begin{enumerate}
\item
as an one-step event, when an individual molecule 
merges with an unoccupied site of the surface of a silicate sheet,

\item
as a two-step event, when a water molecule (i) separates
from a cluster, and (ii) merges with an appropriate site.
\end{enumerate}
At small concentrations of clay, when diffusivity, $D$, is rather
large and clustering of water molecules is negligible,
immobilization of small molecules takes place by the 
one-step mechanism, which means that the rate of 
bounding of penetrant molecules is relatively high.
With an increase in the clay content (which is tantamount 
to a decrease in the average diffusivity, $D$, see Figure 11),
the one-step mechanism of bounding is replaced by the 
two-step mechanism.
According to the proposed scenario, these changes 
in the immobilization mechanism result in a noticeable deceleration
of the process of water bounding to silica sheets,
because separation of a water molecule from a cluster
requires an additional energy provided by thermal fluctuations.
This slowing down (reflected by the model as a decrease in
the coefficient $K$) is confirmed by the experimental data
presented in Figure 13.
This figure demonstrates that $K$ is reduced by at least two
orders of magnitude with the growth of clay content.

It is natural to assume that the total moisture uptake 
by the clay particles, $R_{\rm clay}$, is proportional to 
their concentration, $\nu$.
The parameter $R_{\rm clay}$ is plotted in Figure 15 
versus clay content.
The experimental data show that the linear dependence
(27) is in good agreement (within the experimental
uncertainties) with the observations.
This conclusion may be treated as a confirmation of our kinetic 
equations for anomalous moisture diffusion.

Another confirmation of the model is provided by the fact
that the diffusivity, $D$, becomes practically independent
of the clay content when $\nu$ exceeds 5 wt.-\% (Figure 11).
This conclusion is in accord with numerous observations
that reveal substantial improvement of material properties
of hybrid nanocomposites with clay content in the range of several 
(conventionally, 5) wt.-\%,
whereas at higher clay fractions, material parameters are
weakly affected by the MMT concentration.

\section{Concluding remarks}

Diffusion tests and uniaxial tensile tests with a constant strain rate
have been performed on an intercalated nanocomposite 
with vinyl ester resin matrix and montmorillonite clay filler 
at room temperature.
Observations in diffusion tests show that moisture diffusion in
the neat resin is Fickian, whereas it becomes noticeably 
anomalous (non-Fickian) with the growth of the clay content.
This transition is attributed to immobilization of the penetrant 
molecules on hydrophilic surfaces of clay layers.
Experimental data in mechanical tests demonstrate that the 
response of the neat vinyl ester resin is strongly elastoplastic,
whereas plastic strains strongly decrease with the clay content.
This observations are ascribed to slowing down of molecular
mobility in the host matrix driven by confinement of chains
in galleries between clay layers.

Constitutive equations have been developed for
anomalous moisture diffusion through 
and for the elastoplastic behavior of a nanocomposite.
These relations are determined by 3 adjustable parameters 
(for each process) that are found by fitting the experimental data.
Fair agreement is demonstrated between the observations and
the results of numerical simulation.

The following conclusions are drawn:
\begin{enumerate}
\item
The diffusivity, $D$, and the rate of developed plastic flow, $a$,
decrease with clay content following similar dependences.
The effect of clay concentration, $\nu$, on these quantities 
may be explained by slowing down of molecular mobility in the 
polymeric matrix.

\item
The moisture uptake by the host matrix grows with clay 
content.
With reference to the free volume concept, this observation
is explained by clustering of water molecules in the close 
vicinity of stacks of platelets (where diffusivity falls down 
dramatically).

\item
The rate of immobilization of water molecules on the 
 hydrophilic surfaces of inorganic sheets decreases 
noticeably with $\nu$,
which is associated with transition from the one-step to the
two-step mechanism of immobilization.
\end{enumerate}

An important advantage of our model compared to the tortuous path
concept is that it can explain (in a unified manner) a decay in
diffusivity of a hybrid nanocomposite (with the clay content) 
together with an increase in the maximal moisture 
uptake by the polymeric matrix and a reduction of its
elastoplastic properties.

It is worth noting that the correlations revealed between
a decrease in diffusivity and a decrease in the rate of a steady
plastic flow provide a way to substantially reduce the number 
of (time-consuming) diffusion tests by replacing
a part of experiments by mechanical tests that do not
require long-time observations.

\newpage
\section*{List of figures}
\parindent 0mm

{\bf Figure 1:}
The relative water uptake $\Phi$ versus
the reduced time $\bar{t}$ (h$^{\frac{1}{2}}$/mm).
Circles: experimental data ($2l=0.49953$ mm, $\nu=0.0$ wt.-\%).
Solid line: results of numerical simulation
\vspace*{1 mm}

{\bf Figure 2:}
The relative water uptake $\Phi$ versus
the reduced time $\bar{t}$ (h$^{\frac{1}{2}}$/mm).
Circles: experimental data ($2l=0.17318$ mm, $\nu=0.5$ wt.-\%).
Solid line: results of numerical simulation
\vspace*{1 mm}

{\bf Figure 3:}
The relative water uptake $\Phi$ versus
the reduced time $\bar{t}$ (h$^{\frac{1}{2}}$/mm).
Circles: experimental data ($2l=0.18344$ mm, $\nu=1.0$ wt.-\%).
Solid line: results of numerical simulation
\vspace*{1 mm}

{\bf Figure 4:}
The relative water uptake $\Phi$ versus
the reduced time $\bar{t}$ (h$^{\frac{1}{2}}$/mm).
Circles: experimental data ($2l=0.1397$ mm, $\nu=2.5$ wt.-\%).
Solid line: results of numerical simulation
\vspace*{1 mm}

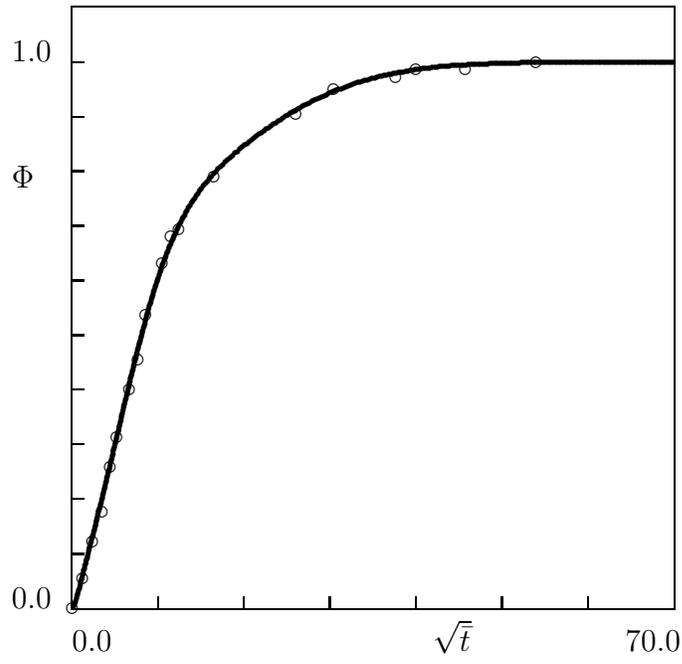
{\bf Figure 5:}
The relative water uptake $\Phi$ versus
the reduced time $\bar{t}$ (h$^{\frac{1}{2}}$/mm).
Circles: experimental data ($2l=0.18452$ mm, $\nu=5.0$ wt.-\%).
Solid line: results of numerical simulation
\vspace*{1 mm}

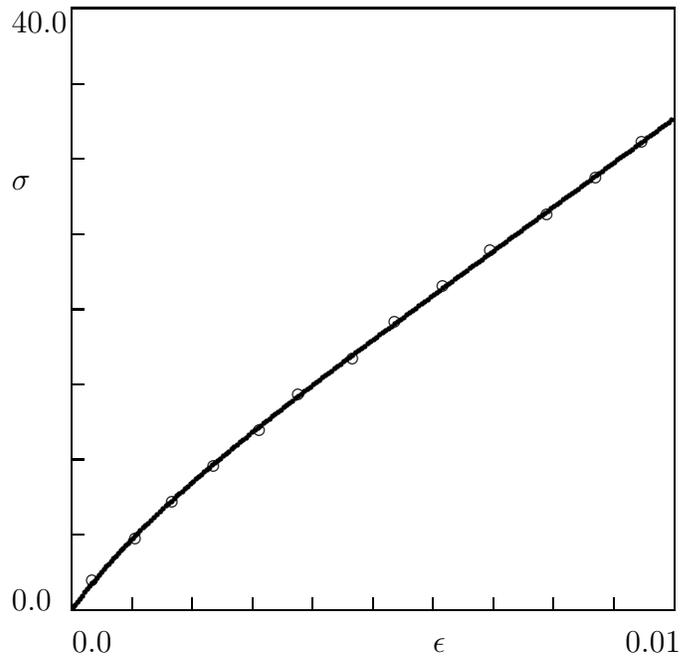
{\bf Figure 6:}
The stress $\sigma$ MPa versus strain $\epsilon$.
Circles: experimental data ($\nu=0.0$ wt.-\%).
Solid line: results of numerical simulation
\vspace*{1 mm}

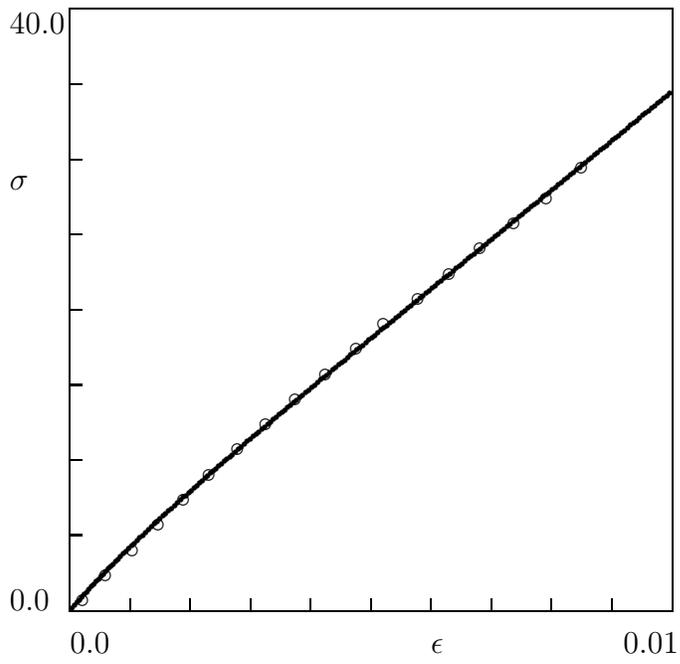
{\bf Figure 7:}
The stress $\sigma$ MPa versus strain $\epsilon$.
Circles: experimental data ($\nu=0.5$ wt.-\%).
Solid line: results of numerical simulation
\vspace*{1 mm}

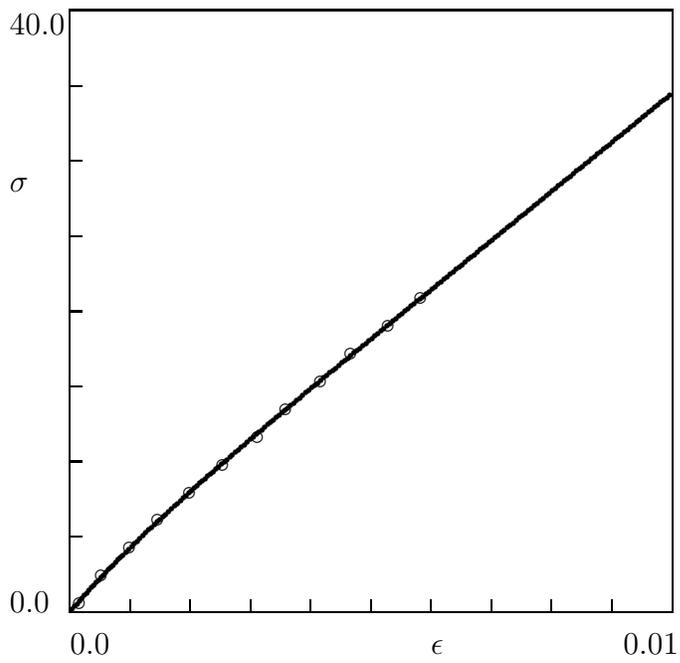
{\bf Figure 8:}
The stress $\sigma$ MPa versus strain $\epsilon$.
Circles: experimental data ($\nu=1.0$ wt.-\%).
Solid line: results of numerical simulation
\vspace*{1 mm}

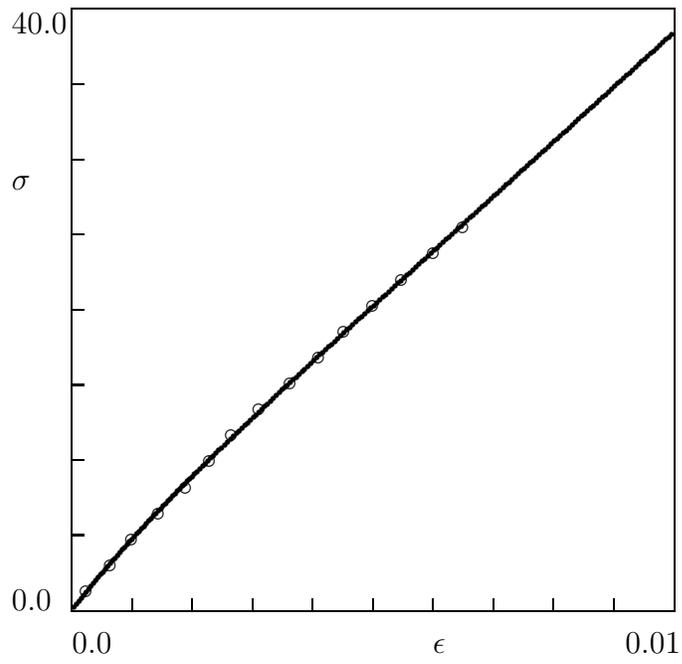
{\bf Figure 9:}
The stress $\sigma$ MPa versus strain $\epsilon$.
Circles: experimental data ($\nu=2.5$ wt.-\%).
Solid line: results of numerical simulation
\vspace*{1 mm}

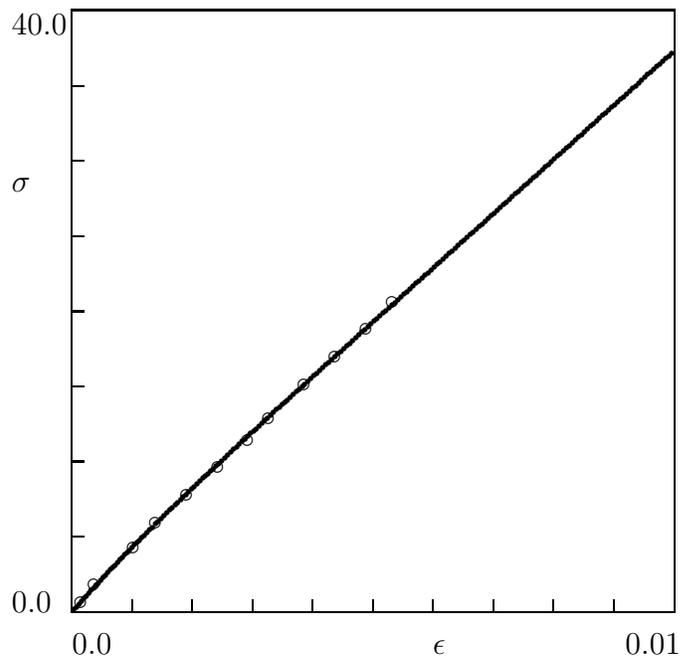
{\bf Figure 10:}
The stress $\sigma$ MPa versus strain $\epsilon$.
Circles: experimental data ($\nu=5.0$ wt.-\%).
Solid line: results of numerical simulation
\vspace*{1 mm}

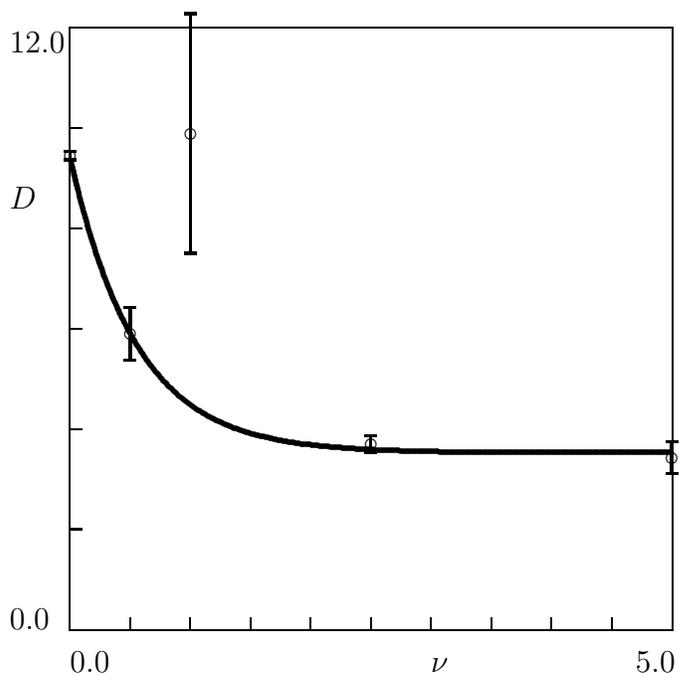
{\bf Figure 11:}
Diffusivity $D\cdot 10^{-7}$ mm$^{2}$/s
versus the clay content $\nu$ wt.-\%.
Circles: treatment of observations.
Solid line: approximation of the experimental data
by Eq. (22) with $D_{0}=3.53$ and $D_{1}=5.93$
\vspace*{1 mm}

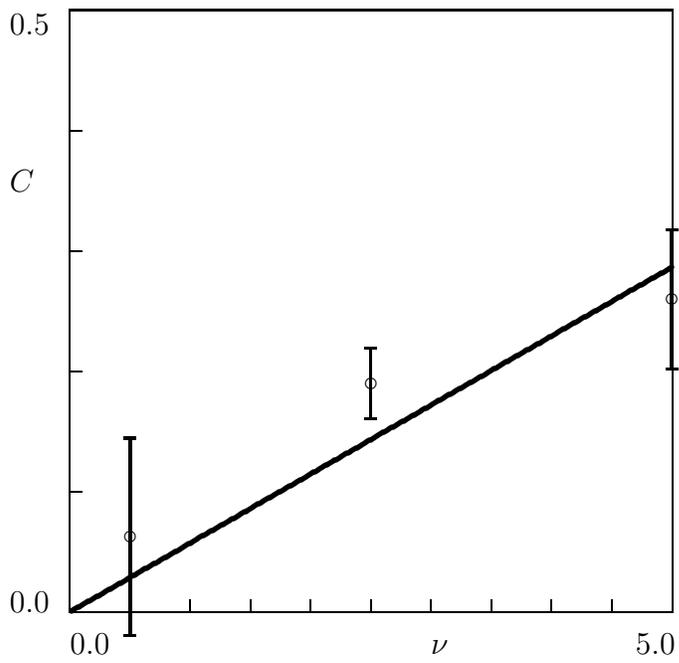
{\bf Figure 12:}
The dimensionless parameter $C$ 
versus the clay content $\nu$ wt.-\%.
Circles: treatment of observations.
Solid line: approximation of the experimental data
by Eq. (23) with $C_{1}=0.057$
\vspace*{1 mm}

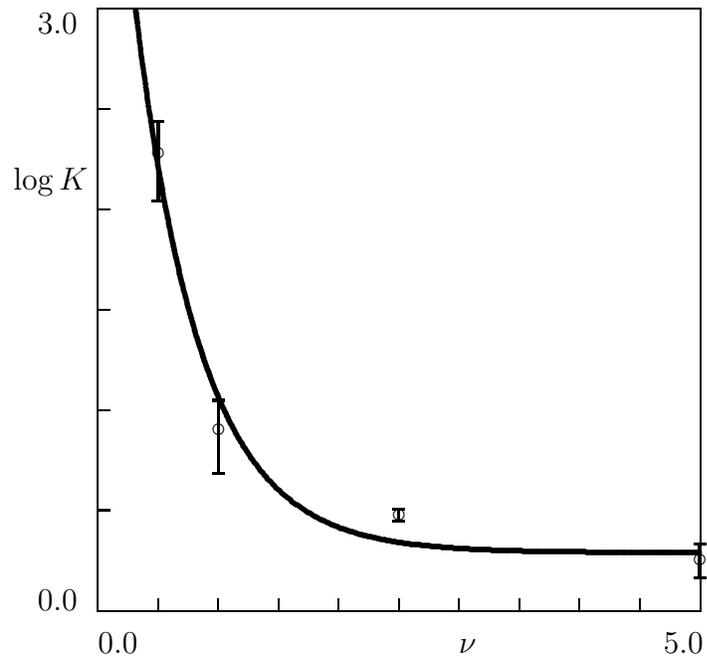
{\bf Figure 13:}
The rate of adhesion $K$ versus the clay content $\nu$ wt.-\%.
Circles: treatment of observations.
Solid line: approximation of the experimental data
by Eq. (24) with $K_{1}=0.29$ and $K_{1}=4.80$
\vspace*{1 mm}

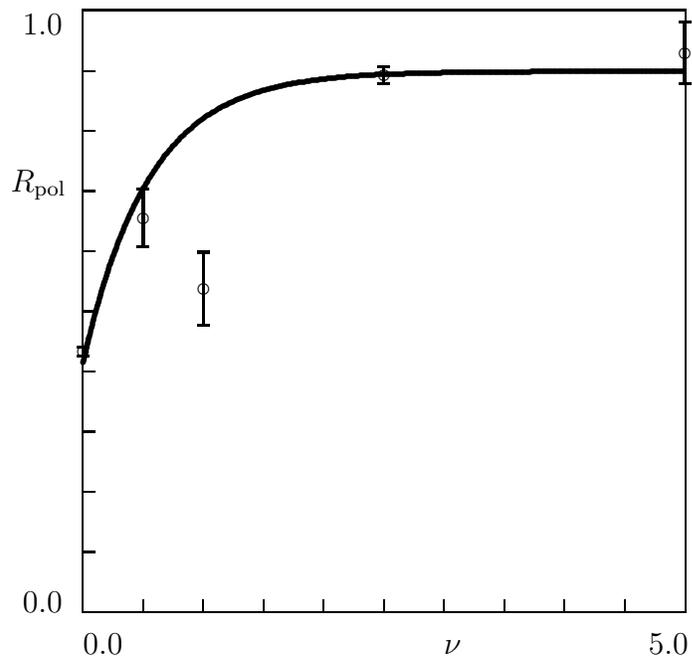
{\bf Figure 14:}
The moisture adsorption by the polymer $R_{\rm pol}$ \%
versus the clay content $\nu$ wt.-\%.
Circles: treatment of observations.
Solid line: approximation of the experimental data
by Eq. (26) with $R_{0}=0.90$ and $R_{1}=0.49$
\vspace*{1 mm}

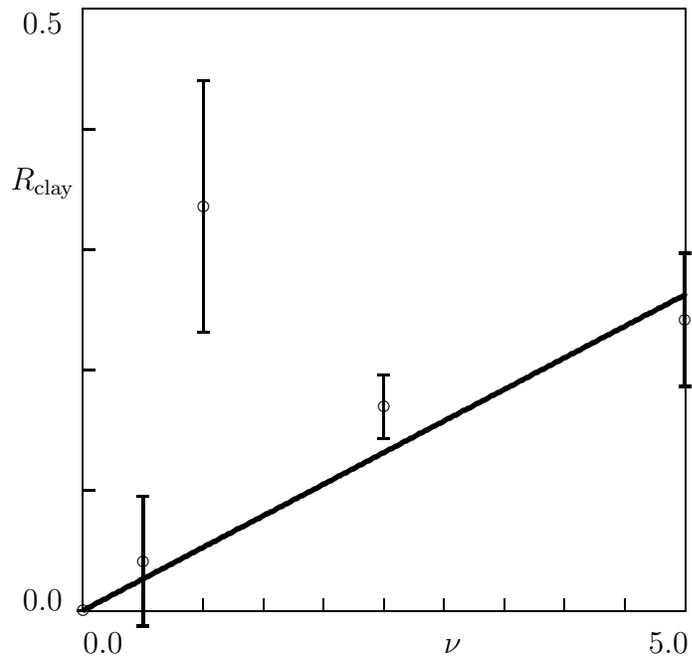
{\bf Figure 15:}
The maximal moisture uptake by the clay particles
$R_{\rm clay}=\kappa n_{1}^{\circ}/\rho$ \%
versus the clay content $\nu$ wt.-\%.
Circles: treatment of observations.
Solid line: approximation of the experimental data
by Eq. (27) with $R_{1}=0.053$
\vspace*{1 mm}

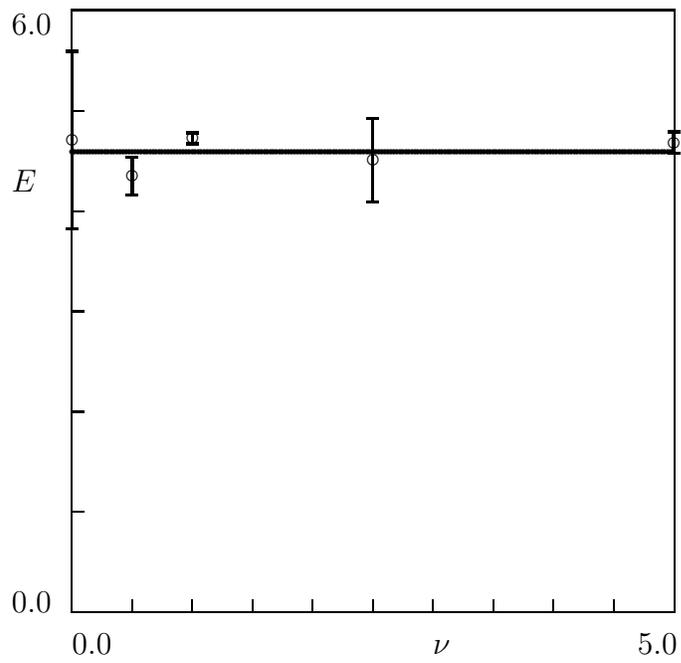
{\bf Figure 16:}
The elastic modulus $E$ GPa 
versus the clay concentration $\nu$ wt.-\%.
Circles: treatment of observations.
Solid line: approximation of the experimental data
by the constant $E=4.60$ GPa
\vspace*{1 mm}

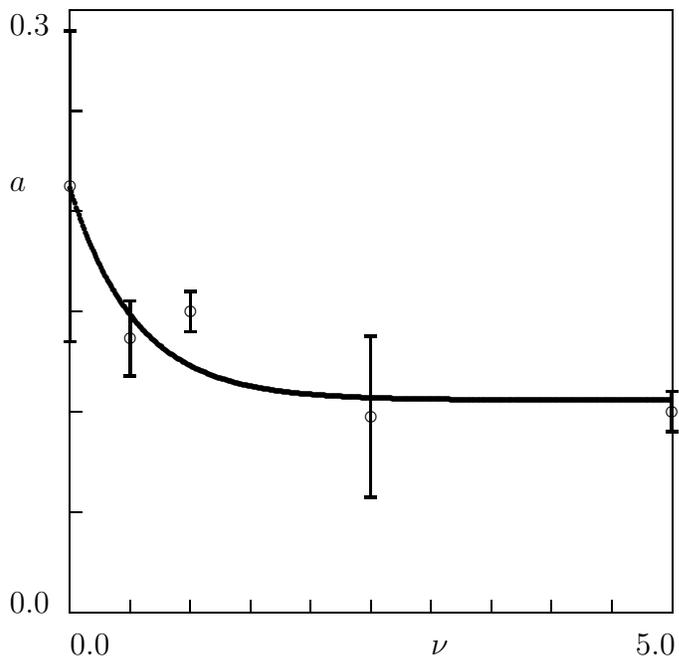
{\bf Figure 17:}
The rate of developed plastic flow $a$ 
versus the clay content $\nu$ wt.-\%.
Circles: treatment of observations.
Solid line: approximation of the experimental data
by Eq. (35) with $a_{0}=0.106$ and $a_{1}=0.106$

\vspace*{100 mm}

\setlength{\unitlength}{0.8 mm}
\begin{figure}[tbh]
\begin{center}

\end{center}
\vspace*{10 mm}

\caption{}
\end{figure}

\begin{figure}[tbh]
\begin{center}
%
\end{center}
\vspace*{10 mm}

\caption{}
\end{figure}

\begin{figure}[tbh]
\begin{center}
%
\end{center}
\vspace*{10 mm}

\caption{}
\end{figure}

\begin{figure}[tbh]
\begin{center}
%
\end{center}
\vspace*{10 mm}

\caption{}
\end{figure}

\end{document}